\documentclass[twocolumn,preprintnumbers,superscriptaddress,amsmath,amssymb]{revtex4-2}
\usepackage{graphicx,dcolumn,bm} 
\usepackage{mathrsfs,tensor}
\usepackage[colorlinks=true
,urlcolor=blue
,anchorcolor=blue
,citecolor=magenta
,filecolor=blue
,linkcolor=blue
,menucolor=blue
,linktocpage=true
,pdfa=true]{hyperref}
\usepackage[dvipsnames,usenames]{xcolor}
\usepackage{enumitem}
\usepackage{soul}
\usepackage{tikz}
\usepackage{tikz-feynman}
\usetikzlibrary{3d,calc}

\newcommand{\Lag}{\mathscr{L}}
\newcommand{\cM}{\mathcal{M}}
\newcommand{\phic}{\phi_{\mathrm{cl}}}

\newcommand{\de}{\partial}
\newcommand{\ie}{\emph{i.e.}}
\newcommand{\eg}{\emph{e.g.}}

\begin{document}

\preprint{COMETA-2025-38 }
\preprint{CERN-TH-2025-187}

\title{Scalar Amplitudes from Fibre Bundle Geometry}
\author{Mohammad Alminawi}
\email{mohammad.alminawi@physik.uzh.ch}
\affiliation{Physik Institut, Universit\"at Z\"urich, Switzerland}
\author{Ilaria Brivio}%
\email{ilaria.brivio@unibo.it}
\affiliation{
Dipartimento di Fisica e Astronomia, Universit\`a di Bologna and INFN, Sezione di Bologna, Italy
}
\affiliation{Physik Institut, Universit\"at Z\"urich, Switzerland}

\author{Joe Davighi}
\email{joseph.davighi@cern.ch}
\affiliation{
Theoretical Physics Department, CERN, Switzerland
}%

\date{\today}

\begin{abstract}
We compute tree-level $n$-point scattering amplitudes in scalar field theories in terms of geometric invariants on a fibre bundle. 
All 0- and 2-derivative interactions are incorporated into a metric on this bundle.
The on-shell amplitudes can be efficiently pieced together from covariant Feynman rules, and we present a general closed formula for obtaining the $n$-point amplitude in this way.
The covariant Feynman rules themselves can be derived using a generalization of the normal coordinate expansion of the fibre bundle metric.
We demonstrate the efficiency of this approach by computing the covariant Feynman rules up to $n=10$ points, from which one can obtain the full amplitudes using our general formula.
The formalism offers a prototype for obtaining geometric amplitudes in theories with {\em higher}-derivative interactions, by passing from the fibre bundle to its jet bundles. 
\end{abstract}

                           
\maketitle
\section{Introduction}
The use of geometrical methods to characterize quantum field theories (QFTs) of scalar fields $\phi^i$ is rooted in the observation that the kinetic terms plus interactions with exactly two spacetime derivatives can 
be cast in a universal form $\Lag \supset \frac{1}{2}\de_\mu\phi^i\de^\mu\phi^j g_{ij}(\phi)$, that defines a  
metric with components $g_{ij}$ on a Riemannian manifold $\mathcal{M}$~\cite{Meetz:1969as,Ecker:1971xko,Vilkovisky:1984st}.
A key insight of this `field space geometry' formalism is that on-shell scattering amplitudes can be expressed using geometric objects which are covariant 
under non-derivative field redefinitions. This feature makes it attractive
for the characterization of theories with non-linear field representations~\cite{Honerkamp:1971sh,Tataru:1975ys,Alvarez-Gaume:1981exa,Alvarez-Gaume:1981exv,Gaillard:1985uh}, 
notably
the Higgs Effective Field Theory (HEFT)~\cite{Alonso:2015fsp,Alonso:2016oah,Cohen:2020xca,Cohen:2021ucp,Alonso:2021rac,Alonso:2023upf}. This has triggered a revival of geometric techniques, that have been extended to gauge and fermion fields~\cite{Finn:2020nvn,Helset:2022tlf,Assi:2023zid,Gattus:2024ird,Assi:2025fsm,Craig:2025uoc}, applied to the derivation of soft theorems~\cite{Cheung:2021yog,Cheung:2022vnd,Derda:2024jvo,Cohen:2025dex}, to EFT renormalization~\cite{Alonso:2022ffe,Helset:2022pde,Jenkins:2023bls,Jenkins:2023rtg,Aigner:2025xyt} and EFT matching~\cite{Li:2024ciy}.

Arguably, this field space geometry approach suffers from two main limitations: (i) interactions with more (or fewer) than two derivatives are not assigned a geometric interpretation, and (ii) covariance is not manifest under derivative field redefinitions. Overcoming these two challenges would be a big step forward in applying geometrical methods to truly general EFTs, and would solidify the interpretation of geometric objects (such as Riemann tensors and their derivatives) as measurable quantities. 
Several resolutions have been recently put forth, notably the functional geometry approach which introduces a momentum-dependent metric~\cite{Finn:2019aip,Cheung:2022vnd,Helset:2022tlf,Cohen:2022uuw,Cohen:2023ekv,Cohen:2024bml,Cohen:2025toappear}. Other approaches modify the fundamental field space, considering Lagrange spaces~\cite{Craig:2023wni} or jet bundles~\cite{Craig:2023hhp,Alminawi:2023qtf}.

In this work we extend the jet bundle formalism we proposed in Ref.~\cite{Alminawi:2023qtf}, 
by applying it to the concrete computation of scattering amplitudes. 
In fact, to make progress in this direction we retreat from the use of metrics on the $r^{\mathrm{th}}$-jet bundle (for some $r\in \mathbb{Z}_{\geq 0}$), which was shown to capture all scalar EFT Lagrangians with up to $2(r+1)$-derivatives~\cite{Alminawi:2023qtf}, 
and focus on the simplest case of the `0-jet bundle'. 
In this case, the scalar field is a section of a fibre bundle whose base is Minkowski spacetime $\Sigma$ and whose fibre $F \cong \mathcal{M}$. By passing from $\mathcal{M}$ to the bundle, we keep track of the spacetime dependence of the field, 
and we capture both the scalar potential and the full 2-derivative action via a metric on this bundle.

Our main goal in this Letter is to showcase the power of this formalism, even at the 0-jet order, by presenting new expressions for general $n$-point scattering amplitudes as functions of geometric tensors on the fibre bundle (\S\ref{sec.results}). 
We show how tree-level on-shell amplitudes can be efficiently computed via covariant Feynman rules, an old notion~\cite{Honerkamp:1971sh,Ecker:1972tii,Honerkamp:1974bp} that was recently revived in~\cite{Finn:2019aip}, removing significant computational hurdles associated to the non-tensorial components of traditional Feynman rules. 
These covariant Feynman rules can be efficiently derived in our formalism from the Taylor expansion of the fibre bundle metric.
We push our explicit results to $n=10$ points, and present a new formula for obtaining the full $n$-point amplitude from its covariant building blocks~\cite{Alminawi:future}. 

In future work, we will extend this formalism beyond tree-level, and to higher $r$-jet order. In that setting, the entire EFT basis up to arbitrary fixed order in the derivative expansion is captured by a fibre bundle metric~\cite{Alminawi:2023qtf}. Thus, we expect the same techniques we develop here, for passing from bundle metric to amplitudes, should deliver geometric formulae for amplitudes in theories with higher derivatives. The present paper lays important technical groundwork towards achieving this goal.

\section{Formalism}\label{sec.formalism}

In this Section we review general properties of scattering amplitudes in scalar theories, taking care to highlight  covariance properties. This motivates our fibre bundle formalism for describing theories with 0- and 2-derivative interactions; the fact that all couplings are subsumed consistently into a single metric tensor provides an efficient formalism for building amplitudes out of all appropriate covariants that we can derive from that metric.

\subsection{Amplitudes and their covariance}\label{sec.covariance}

Consider the quantum dynamics of a set of scalar fields $\phi^i(x), \,i=1\dots N$, that we assume can be specified by an action functional $S[\phi^i(x)]$. 
For an $n$-particle scattering $\phi^{a_1}(p_1)... \phi^{a_m}(p_m)\;\to\; \phi^{a_{m+1}}(-p_{m+1})... \phi^{a_n}(-p_n)$,
where $a_i$ are fixed flavour indices valued in $1\dots N$ and $p_i$ are the ingoing four-momenta,
the $\mathcal{S}$-matrix element is given by the Lehmann--Symanzik--Zimmermann (LSZ) formula,
\begin{align}
\label{eq.LSZ}
\langle p_n... p_{m+1}|i\mathcal{T} |p_1... p_m\rangle_{a_1... a_n} = (2\pi)^4\delta^{(4)}({\textstyle\sum} p_i)\,\mathcal{A}_{a_1... a_n} 
\end{align}
where $\mathcal{S}=1-i\mathcal{T}$ and $\mathcal{A}_{a_1\dots a_n}$ is the connected, amputated $n$-point function. 
The standard procedure to calculate $\mathcal{A}_{a_1...a_n}$ is to first obtain the effective action 
$\Gamma[\phi]=S[\phi] + (i/2)\log \det \left( -{\delta^2 S}/{\delta\phi\delta\phi} \right) + \dots$ 
From there, the one-particle-irreducible (1PI) $k$-point functions are the functional derivatives
$\delta^k (i\Gamma)/\delta \phi^{i_1}(x_1)... \delta \phi^{i_k}(x_k)|_{\phi=\phi_{\text{cl}}}$,
where ${\phic=\langle 0 |\phi|0\rangle}$ is the classical field configuration defined by the vacuum condition
$\delta \Gamma/\delta \phi|_{\phi=\phic}=0$.
Of particular importance, at $k=2$ we have that
\begin{equation} \label{eq:propagator-1}
\frac{\delta^2(i\Gamma)}{\delta\phi^{i}(x_i)\delta\phi^{j}(x_j)}\bigg|_{\phi=\phic} =: - (D_F)^{-1}_{ij}(x_i-x_j)
\end{equation}
defines the inverse Feynman propagator in position space. 
Under a field redefinition $\phi^i(x) \mapsto \psi^i(x)$, twice applying the chain rule tells us
\begin{equation} \label{eq:propagator}
    \frac{i\delta^2\Gamma[\psi]}{\delta\psi^{a}\delta\psi^{b}}\bigg|_{\psi_{\mathrm{cl}}} = \tensor{U}{^{i}_{a}}\tensor{U}{^{j}_{b}}
    \frac{i\delta^2\Gamma[\phi]}{\delta\phi^{i}\delta\phi^{j}}\bigg|_{\phic}, \quad 
    \tensor{U}{^i_a} := \frac{\delta \phi^a}{\delta \psi^i}\bigg|_{\psi_{\mathrm{cl}}},
\end{equation}
iff we impose the vacuum condition ${\delta\Gamma/\delta\psi|_{\psi=\psi_{\mathrm{cl}}}=0}$~\footnote{If the vacuum is $\langle\phi^i\rangle\equiv 0\equiv\langle\psi^a \rangle$, $U^i_a$ selects the pure linear term in the field transformation. Indeed field redefinitions that leave both the vacuum and the mass spectrum invariant are of the type $\phi^i\to U^i_a\psi^a+\dots$}.
Thus the propagator transforms as a symmetric tensor under field redefinitions, up to the vacuum condition~\cite{Cohen:2023ekv}.

Starting from propagators and the 1PI $k$-point functions, the connected $n$-point functions $\mathcal{A}_{a_1...a_n}$ are then built by Wick-contracting 1PI $k$-point functions (with $3\leq k \leq n$) via $D_F$ internal lines, forming all possible tree-level topologies that make diagrams with $n$ external legs. 
Unlike inverse propagators, IP1 $k$-point functions with $k\geq 3$ are not tensors. However, the amputated connected functions $\mathcal{A}_{a_1...a_n}$ 
do transform as tensors of type $(0,n)$ under field redefinitions, \ie\
\begin{equation}
\label{eq.A_tensor}
\mathcal{A}_{a_1... a_n}
= \tensor{U}{^{i_1}_{a_1}}\cdots \tensor{U}{^{i_n}_{a_n}} \mathcal{A}_{i_1... i_n}\,\, ,
\end{equation}
provided we impose (i) the vacuum condition ${\delta\Gamma/\delta\psi|_{\psi_{\mathrm{cl}}}=0}$, and (ii) the `on-shell condition' for all the external legs, loosely that (see \S \ref{sec:2-pt})
\begin{equation}
    \frac{\delta^2 \Gamma[\psi]}{\delta \psi \delta \psi}\bigg|_{\psi=\psi_{\mathrm{cl}},\, p^2=m^2}=0\, .
\end{equation}
This highly non-trivial property follows from an intricate set of cancellations of the non-tensorial parts between different Feynman graphs. 
As will be shown in~\cite{Alminawi:future}, and in accordance with the recursive argument of~\cite{Cohen:2023ekv}, these cancellations occur at arbitrary finite $n$ essentially because the Feynman graph expansion -- itself a property of the path integral -- is a nested diagrammatic representation of the Fa\`a di Bruno formula for applying the chain rule $k$ times: 
\begin{equation} \label{eq:FdB}
    \frac{\delta^k}{\delta\phi^k}\Gamma(\psi(\phi)) = \sum_{l=0}^k \Gamma^{(l)}(\psi)\cdot B_{k,l}(\psi^{(1)},\dots,\psi^{(k-l+1)}),
\end{equation}
where $f^{(n)}(x)\equiv \delta^n f(x)/\delta x^n$, and
$B_{n,k}$ are the partial exponential Bell polynomials.
In the Supplemental Material we illustrate how this works in the stripped-down context of a 0-d scalar theory. 
Inserting~\eqref{eq.A_tensor} into the LSZ formula~\eqref{eq.LSZ} ensures that, for fixed indices $a_1...a_n$ \ie\ fixed physical external states, 
the matrix element is \emph{in}variant under field redefinitions -- a textbook result in QFT~\cite{Chisholm:1961tha,Kamefuchi:1961sb}.

The same combinatoric technology set out in~\cite{Alminawi:future}
can be used to establish the important corollaries:
\begin{itemize}[leftmargin=*]
    \item That the on-shell, tree-level amplitudes can be equivalently derived using {\em covariant Feynman rules}, as introduced in~\cite{Honerkamp:1971sh,Ecker:1972tii,Honerkamp:1974bp,Finn:2019aip}, that coincide with canonical Feynman rules up to non-tensorial terms;
    \item That the on-shell $n$-point amplitude $\mathcal{A}_{a_1\dots a_n}$, henceforth abbreviated to $\mathcal{A}_n$, can be obtained from these covariant Feynman rules via the following closed formula, that is to our knowledge new~\cite{Alminawi:future}:
    \begin{align} \label{eq:closed}
        \mathcal{A}_{n}=\sum_{k=1}^{n-2} &\,\frac{(n+k-2)!}{(n-2)!} \Delta^{1-k} \\
        &\times B_{n-2,k}\left(\frac{1}{2}\mathcal{R}_{3},\dots, \frac{1}{n-k}\mathcal{R}_{n-k+1} \right) \nonumber
    \end{align}
    where $\Delta$ is the momentum space propagator (see \S\ref{sec:2-pt}), $B_{n,k}$ are the same partial Bell polynomials appearing in the Fa\`a di Bruno formula above, and $\mathcal{R}_{l}$ denotes the $l$-point covariant Feynman rule, whose $l$ indices (and how they are contracted with the propagator indices) we have suppressed for brevity. A symmetrization over these indices is implied. Special cases of this general formula have been discussed in {\em e.g.}~\cite{Kreimer:2016jxo,Balduf:2019mai}. 
\end{itemize}
A central result of the present paper (see \S \ref{sec.results}) is to reveal the structure of the covariant Feynman rules $\mathcal{R}^{(l)}$ for scalar theories with 0- and 2-derivative terms, and to efficiently put them together into amplitudes with many external legs using~\eqref{eq:closed}.

\subsection{Geometric theories via fibre bundles}

The statements so far regarding covariance hold for general QFTs. They become especially powerful for theories in which we can systematically build objects that are covariant under (possibly a subset of) field redefinitions. This is where `geometrical methods' enter the game. In short, if the couplings of a QFT can be packaged into a small number of tensors on some auxiliary target space, then we can hope to build covariant objects (under field redefinitions) out of tensors on that target space, and thence build amplitudes. 
A theory of Goldstone bosons admits such a treatment: at leading order in the derivative expansion, the relevant interactions feature 2 spacetime derivatives and can be obtained from a metric on a Riemannian manifold $\cM$. The amplitudes are built from the associated Riemann tensor and covariant derivatives thereof~\cite{Alonso:2015fsp,Alonso:2016oah,Cheung:2021yog,Cheung:2022vnd,Derda:2024jvo,Cohen:2025dex}. 

\begin{figure}
    \centering
    \begin{tikzpicture}[scale=1.2,
    every node/.style={font=\small},
    axis/.style={->, thick},
    fiber/.style={thick},
    vertvec/.style={blue, thick},
    horizvec/.style={red}
]

% Base space X
\draw[thick] (-2,-0.8) .. controls (-1,-1.0) and (1,-1.0) .. (2,-0.8)
             .. controls (1.5,-0.6) and (-1.5,-0.6) .. cycle;
\node at (-1.8,-1.0) {$\Sigma$};

% Projection arrows to base
\draw[dotted,thick] (-1,0.6) -- (-1,-0.88) node[below] {$x$};

% Total space P (drawn as curved surface)
\draw[thin] (-2,0) .. controls (-1,0.3) and (1,0.3) .. (2,0.1)
             .. controls (1.5,-0.3) and (-1.5,-0.3) .. cycle;
\draw[thin] (-2,1.5) .. controls (-1,1.8) and (1,1.8) .. (2,1.6)
             .. controls (1.5,1.3) and (-1.5,1.3) .. cycle;
\draw[thin] (-2,1.5) -- (-2,0);
\draw[thin] (2,1.6) -- (2,0.1);

% Fibers Px and Py
\draw[fiber] (-1,-0.1) .. controls (-1.4,0.0) and (-0.6, 1.2) .. (-1,1.55) node[right] {{\footnotesize $\pi^{-1}(x) \cong \cM$}};

% Projection arrow π
\draw[axis,thin] (-2.5,1.4) -- (-2.5,-0.7) node[midway,left] {$\pi$};

% Points p and q in P
\fill (-1,0.6) circle(1.5pt) node[above left] {$e$};

% Vertical vectors Vp and Vq
\draw[vertvec] (-1.22,0.0) -- (-0.78,1.2) node[below right] {$V_e$};

% Horizontal spaces Hp and Hq
\draw[horizvec] (-1.35,0.5) -- ++(0.5,0.3) -- ++(0.2,-0.1) -- ++(-0.5,-0.3) -- cycle
    node[pos=0.5,right] {$\,\,\,\quad H_e$};

% section passing through both points
\draw[magenta, thick, dashed] (-2,0.4) .. controls (0,0.95) and (1,0.4) .. (2,0.6) node[right] {$\phi$};

\end{tikzpicture}
    \caption{Illustration of the fibre bundle geometry $(E,\pi,\Sigma)$, with a section $\phi$ labelled that passes through a point $e\in \pi^{-1}(x)$ in the bundle, at which the vertical space $V_e$ and a choice of horizontal space $H_e$ are also sketched.}
    \label{fig:Ve}
\end{figure}
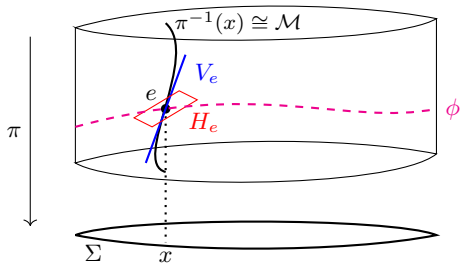

In~\cite{Alminawi:2023qtf} we showed that, via a generalisation of this formalism, one can incorporate 0-derivative interactions (and masses) into a target geometry. 
To wit, we describe $N$ scalar fields $\phi^i(x)$ using a fibre bundle $(E,\Sigma,\pi)$; here $E$ is the total space, $\Sigma$ the base space which we take to be flat Minkowski spacetime, and $\pi:E\to \Sigma$ is a surjective submersion called the projection map (see Fig.~\ref{fig:Ve}). 
For an open neighbourhood $\mathcal{U}_x \subset \Sigma$ containing a point $x\in\Sigma$, there exists a local trivialisation $\varphi:\pi^{-1}(\mathcal{U}_x)\mapsto \mathcal{U}_x \times \cM$, where $\cM$ is itself a manifold called the {\em fibre}, which is the target space of our scalar fields. The dimension of $\cM$ equals the number of real scalar degrees of freedom.
We introduce local fibred coordinates $(x^\mu, u^i)$ on $\pi^{-1}(\mathcal{U}_x)$, with $x^\mu$ being coordinates on the base $\Sigma$ and $u^i$ coordinatizing $\cM$.

A field configuration is a section of this bundle, that is a smooth map $\phi:\Sigma \to E$ such that $\pi \circ \phi = \text{id}_\Sigma$.
We let $\Gamma(\pi)$ denote the set of all sections of $\pi$. It is the (infinite-dimensional) space of all smooth field configurations. 
The arguments of \S \ref{sec.covariance} imply the amplitudes $\mathcal{A}_{a_1\dots a_n}$ ought to be tensors on $\Gamma(\pi)$, up to imposing the vacuum and on-shell conditions.

To build a Lagrangian, we equip the fibre bundle $(E,\Sigma,\pi)$ with a pseudo-Riemannian metric $g$. In our fibred coordinate system, $g$ takes the block diagonal form:
\begin{equation}
g = g_{\mu\nu}(u) \,dx^\mu\otimes dx^\nu +g_{ij}(u)\, du^i\otimes du^j\, .
\end{equation}
Poincar\'e invariance has been used to (i) set the mixed components $\sim dx^\mu\otimes du^i$ vanish, and (ii) enforce $\partial_\mu g_{\rho\sigma}=\partial_\mu g_{ij}=0$.
The Lagrangian density, evaluated for a particular field configuration {\em i.e.} section $\phi$, is then obtained by pulling back $g$ from $E$ to spacetime $\Sigma$ along $\phi$ before contracting with the inverse spacetime metric: 
\begin{align} \label{eq:Lag}
\Lag &=
\frac{1}{2}\left\langle\eta^{-1},\phi^*(g)\right\rangle \\
&= 
\frac{\Lambda^4}{2}\eta^{\mu\nu}g_{\mu\nu}(\phi(x)) +
\frac{1}{2}g_{ij}(\phi(x))\,\de_\rho\phi^i(x) \de^\rho\phi^j(x)\,, \nonumber
\end{align}
where $\phi^i := u^i \circ \phi$. Then the action functional is $S[\phi]=\int_\Sigma \Lag \mu_\Sigma$, where $\mu_\Sigma = d^4 x$ is the volume form on $(\Sigma,\eta)$.
Choosing 
$g_{\mu\nu}(u_i) = - \frac{1}{2}\eta_{\mu\nu} V(u_i)$
ensures the first term equals $-\Lambda^4V(\phi_i)$ for some smooth potential function $V$. Eq.~\eqref{eq:Lag} is then the most general scalar EFT Lagrangian with $\leq 2$ derivatives. 
Indeed, we showed in~\cite{Alminawi:2023qtf} that by extending the bundle $E$ to its jet bundles $J^r(E)$, which are a sequence of fibre bundles measuring higher derivatives of sections $\phi$, the entire EFT Lagrangian with up to $2(r+1)$ derivatives can be obtained by pulling back a metric, as in~\eqref{eq:Lag}. Many of the techniques we develop in this paper for passing from a bundle metric to amplitudes should find a home in this higher-derivative setting.

We hereon assume a normalisation $V(0)=1$ for the cosmological constant, 
but our results generalise to other values. (Note, however, that $V(0)\neq 0$ is required for $g$ to be non-singular at $u^i=0$.)
The vacuum condition $\delta \Gamma/\delta\phi|_{\phic}=0$ implies
    $\de_i g_{\mu\nu} = 0$ $\implies \de_i V|_{\phic}=0$,
that $\phic$ should minimise the potential.

\subsection{Amplitude decomposition via vertical tensors}

Now that we have a `geometric theory', with all couplings packaged into a metric $g$ on $E$, we can systematically build covariant objects on $\Gamma(\pi)$ out of tensors on $E$. 
As an illustrative example (not yet related to an amplitude), from $g$ we can define a metric $G$ on $\Gamma(\pi)$ by pointwise evaluation followed by integration over $\Sigma$: declare $G_\phi(\delta_1 \phi, \delta_2 \phi) := \int_\Sigma g_{\phi(x)}(\delta_1 \phi(x), \delta_2 \phi(x)) \mu_\Sigma$ for any pair of tangent vectors $\delta_1 \phi$, $\delta_2 \phi \in T_\phi\Gamma(\pi)$
~\footnote{To see how covariance of $g$ on $E$ implies covariance of $G$ on $\Gamma(\pi)$, consider a change of section $\phi \to \phi^\prime$ that can be captured by a fibre-preserving diffeomorphism $f$ on $E$:
$\phi^\prime = f \circ \phi$, where $f:E \to E \,|\, \pi \circ f = \pi$, {\em a.k.a.} a non-derivative field redefinition~\cite{Alminawi:2023qtf}. Under $f$ the metric is pulled back while the vector fields on which we evaluate it are pushed forward, and $G$ is covariant.}.

Via the same procedure, covariant amplitudes $\mathcal{A}_n$ can be built from tensors on the fibre bundle $E$. Also Taylor expanding in the Mandelstams $s_{ij}=p_i\cdot p_j$ to $\mathcal{O}(s)$ and imposing $\Gamma^{(1)}= \Gamma^{(2)}= 0$, we can write
\begin{align} \label{eq:An}
    \mathcal{A}_n(\delta_1\phi,\dots,\delta_n\phi; s) {=} &\sum_\alpha \int_\Sigma T^{0,\alpha}_{\phi(x)} (\delta_1\phi(x),\dots,\delta_n\phi(x)) \mu_\Sigma \nonumber \\
    +\,\,\sum_{ij}s_{ij}\sum_\alpha \int_\Sigma T^{ij,\alpha}_{\phi(x)} &(\delta_1\phi(x),\dots,\delta_n\phi(x)) \mu_\Sigma 
\end{align}
where each $\alpha$ sums over some tensor structures $T$ on $E$ with the appropriate mass dimension. 
Specifically, Lorentz invariance tells us that it is only {\em vertical tensors} on $E$ that can appear in $\mathcal{A}_n$, which we define in the Supplemental material. Colloquially speaking, vertical tensors only have non-vanishing components in the fibre directions (see Fig.~\ref{fig:Ve}); any $\mu$ indices must be contracted.

Poincar\'e invariance moreover implies $\partial_\mu g_{\rho\sigma}=\partial_\mu g_{ij}=0$,
so we can further restrict to vertical tensors obtained by pulling back tensors on the fibre $\cM$.
The integrals over $\Sigma$ in~\eqref{eq:An} are then redundant, simply averaging over constant functions -- indeed the classical configuration we evaluate on is also constant, meaning we only pick out the value of each vertical tensor at the vacuum point.

Despite restricting to vertical tensors, including the base degrees of freedom in our bundle $E$ allows more geometric invariants to be constructed than if we only had a map to field space and a metric thereon. This is because the Lagrangian~\eqref{eq:Lag} encodes the potential $V$ through its $g_{\mu\nu}$ components: we can build vertical, $x^\mu$-independent tensors from $g$ that nevertheless depend on $V$. 
For example, consider the vertical tensor with components 
\begin{equation} \label{eq:Rij}
    \tilde R_{ij} = R^{\mu}_{\,\,\,i\mu j} = \frac{1}{2\Lambda^4}\mathrm{Tr}[\eta]\partial_{i}\partial_j V\, ,
\end{equation}
\ie\ the mass squared matrix, which will appear in the propagator (see \S \ref{sec.results}). The precise meaning of the index contraction is explained in the Supplemental Material. 
We see from~\eqref{eq:Rij} how, in this formalism, derivatives of $V$ measure curvature components in the fibre bundle, and so capture aspects of the geometry of $E$. In this sense, the potential acquires a geometric meaning.

The sum over such vertical tensors in~\eqref{eq:An} captures all contributions to $\mathcal{A}_n$ from  theories with $\leq 2$ derivatives. The explicit tensors that appear, along with their coefficients, will be presented in \S\ref{sec.results}.
This decomposition dictates, by covariance, the dependence of all $\mathcal{A}_n$ on $V(\phi)$ and its derivatives. 

\section{Tree-level on-shell  amplitudes}\label{sec.results}

We have so far been schematic in describing how amplitudes can be built from vertical tensors on $E$. Now we make this concrete, explaining how covariant Feynman rules offer an efficient tool for deriving $\mathcal{A}_n$ to large $n$.

\subsection{Covariant Feynman rules}
As anticipated in \S\ref{sec.covariance}, tree-level on-shell scattering amplitudes $\mathcal{A}_n$ can be efficiently computed using the general formula~\eqref{eq:closed}. The input we need to evaluate this finite sum is the set of {\em covariant Feynman rules}, that we denote $\mathcal{R}_{a_1\dots a_k}$ (or $\mathcal{R}_k$ for brevity), for each integer $k\in [3,n]$. This notion was revived in~\cite{Finn:2019aip} using functional geometry; in that language, $\mathcal{R}_k$ can be defined as the $k^{\mathrm{th}}$ covariant functional derivative of $\Gamma$.
The $\mathcal{R}_k$ coincide with the usual Feynman rules $F_{a_1\dots a_k}$ up to non-tensor terms $N_{a_1\dots a_k}$:
\begin{equation}
\label{eq.FvsR}
    F_{a_1\dots a_k} (p_{i})= \mathcal{R}_{a_1\dots a_k}(p_i) + N_{a_1\dots a_k}(p_i)\,,
\end{equation}
where $p_i$ represent the external momenta.

Computing the covariant Feynman rules $\mathcal{R}_{a_1\dots a_n}$ becomes especially tractable for a `geometric theory' such as ours.
In the fibre bundle formalism, the $\mathcal{R}_{a_1\dots a_n}$ are vertical tensors on $(E,\Sigma,\pi)$, while  $N_{a_1\dots a_n}$ contain various terms proportional to Christoffel symbols and their derivatives, which can be further decomposed as
\begin{equation}
N_{a_1\dots a_n} = N^{(1)}_{a_1\dots a_n} + N^{(2)}_{a_1\dots a_n} + \bar N_{a_1\dots a_n}\,,
\end{equation}
where $N^{(1)}$ vanishes at the vacuum, $N^{(2)}$ vanishes upon imposing on-shell conditions on all $p_i$, while $\bar N$ remains non-zero.
When computing tree-level on-shell scattering amplitudes $\mathcal{A}_n$ via $F$ insertions, the $\bar N$ terms and the $N^{(2)}$ terms associated to legs contracted in internal propagators cancel amongst each other, so the final result depends only on the $\mathcal{R}_n$ structures as in~\eqref{eq:closed}. We show how these cancellations play out in practice for the case of $2\to 2$ scattering in \S \ref{sec:cov-amplitude} below.

The power of our general formula~\eqref{eq:closed} for amplitudes, which holds independently of the fibre bundle formalism, hinges on how easily the $\mathcal{R}_n$ structures can be identified.  The fibre bundle picture offers a systematic procedure to determine each $\mathcal{R}_n$, including all 0- and 2-derivative contributions, via a single Taylor expansion of the metric on the bundle, the details of which are relegated to App.~\ref{sec.cFR_derivation}.

\subsubsection{Covariant propagators} \label{sec:2-pt}
The Feynman propagator $iD_F^{ij}(x-y)$ in Eq.~\eqref{eq:propagator-1} is the Green's function of the Klein--Gordon equation:
\begin{align}
\label{eq.kleingordon}
(g_{ij} \square_{x}+ \Lambda^4\tensor{R}{^\mu_i_\mu_j}/2)iD_F^{jk}(x-y)   = \delta^k_i\delta^4(x-y) .
\end{align}
As a Feynman rule in momentum space, the tree-level propagator and its inverse are
\begin{align}
\Delta^{a_1a_2}(p^2) &= 
i\left(p^2 g_{a_1a_2} + \Lambda^4 \tensor{R}{^\mu_{a_1}_\mu_{a_2}}/2\right)^{-1} \,,
\\
\Delta^{-1}_{a_1a_2}(p^2) &= 
-i\left(p^2 g_{a_1a_2} + \Lambda^4 \tensor{R}{^\mu_{a_1}_\mu_{a_2}}/2\right) \,,
\label{eq.prop_FR}
\end{align}
in a notation such that $(A_{ij})^{-1}=(A^{-1})^{ij}$ for a (0,2) tensor $A$.
In general $\Delta^{ij}$ is neither diagonal nor canonically normalized, but it is always possible to bring it to a canonical form via a coordinate transformation as in Eq.~\eqref{eq:propagator}.
An on-shell condition can only be defined for mass eigenstates, \ie\ it is $\tensor{U}{_k^i}\tensor{U}{^j_k}\Delta^{-1}_{ij} = 0$ for the asymptotic state $\phi_k$. 
In Eqs.~\eqref{eq.kleingordon}--\eqref{eq.prop_FR} and in all the results of this section, the geometric objects $(g_{a_1a_2},\tensor{R}{^\mu_{a_1}_\mu_{a_2}}\dots)$ are understood to be evaluated at the vacuum. 

\subsubsection{Covariant vertices}

Using the fibre bundle formalism,
we can straightforwardly compute the covariant Feynman rules $\mathcal{R}_n$ to high order, by applying known techniques in differential geometry for Taylor expanding a metric tensor and decomposing the result into tensor and non-tensor pieces~\cite{Muller:1999rnc}. We summarize various simplifying tricks that follow from Poincar\'e invariance in App.~\ref{sec.cFR_derivation}.
We note that the same results could be obtained with the traditional field space formalism by treating the potential $V$ as a separate function and taking covariant derivatives thereof~\cite{Finn:2019aip}. In passing to the bundle, all contributions come from a single metric expansion.
The results up to $n\leq 5$ are:
\begin{align}
\label{eq.R3}
\mathcal{R}_{a_1a_2a_3} &= i\,\frac{\Lambda^4}{2}\nabla_{a_1}\tensor{R}{^\mu_{a_2}_\mu_{a_3}}
\\
\label{eq.R4}
\mathcal{R}_{a_1a_2a_3a_4} &=  \frac{i}{4!} \sum_{S_4}\bigg[
\frac{\Lambda^4}{2}\nabla_{a_1}\nabla_{a_2}\tensor{R}{^\mu_{a_3}_\mu_{a_4}}
\\
&
-2\Lambda^4 \tensor{R}{^\mu_{a_1}_\nu_{a_2}} \tensor{R}{^\nu_{a_3}_\mu_{a_4}}
-2 s_{12} \, R_{a_1a_3a_4a_2}  \bigg]
\nonumber
\\
\label{eq.R5}
\mathcal{R}_{a_1a_2a_3a_4a_5} &= \frac{i}{5!}\sum_{S_5}\bigg[
\frac{\Lambda^4}{2}\nabla_{a_1}\nabla_{a_2}\nabla_{a_3}\tensor{R}{^\mu_{a_4}_\mu_{a_5}}
\\
&
-7 \Lambda^4\tensor{R}{^\mu_{a_1}_\nu_{a_2}} \nabla_{a_3}\tensor{R}{^\nu_{a_4}_\mu_{a_5}}
-5 s_{12} \, \nabla_{a_5}R_{a_1a_4a_3a_2}
\bigg]
\nonumber
\end{align}
where $s_{ij} = (p_i+p_j)^2$, with $p_i$ the incoming 4-momentum corresponding to $a_i$. The sums run over all permutations of the external states $S_n$, acting simultaneously on flavour indices $a_1\dots a_n$ and momenta $p_1\dots p_n$.  Note $\mathcal{R}^{(3)}_{a_1a_2a_3}$ is invariant under such permutations due to the second Bianchi identity. The expressions for vertices with $n\leq 10$ are provided in \S\ref{sec.cFR_derivation}. 

We pause to make several comments. Firstly, we emphasize that both the $s$-dependent and $s$-independent parts of the amplitude appear on an `equal footing', in that both are covariant derivatives of the Riemann tensor on the fibre bundle. The potential contributions, like those from the kinetic term, can be identified with invariants of the bundle geometry. 
Secondly, our covariant Feynman rules in general feature several distinct terms corresponding to $\nabla^n V$, all of which are separately invariant -- in that sense, the bundle provides us with a `finer' set of invariant quantities {\em a.k.a.} observables.

\subsection{Covariant tree-level amplitudes} \label{sec:cov-amplitude}

The covariant Feynman rules can be composed into tree-level amplitudes using the general formula~\eqref{eq:closed} that sums over all Feynman diagrams. As a word of caution, note that the compact form of~\eqref{eq:closed} includes a sum over terms with the same $\mathcal{R}_n$ insertions but different index contractions, that correspond to different Feynman diagrams coming with different multiplicities. The details of the combinatoric coefficients that enter, which we record in Fig.~\ref{fig.A4_A6} for $n=4,\, 5,\, 6$, will be presented in~\cite{Alminawi:future}. 
To illustrate, the first few examples are: 
\allowdisplaybreaks
\begin{align}
\label{eq.A3}
&\mathcal{A}_{a_1a_2a_3} = \mathcal{R}_{a_1a_2a_3}
\\
\label{eq.A4}
&\mathcal{A}_{a_1a_2a_3a_4} = \mathcal{R}_{a_1a_2a_3a_4} \\
&+ \frac{1}{4!}\sum_{S_4} \bigg[3\, \mathcal{R}_{a_1a_2b_1}\Delta^{b_1b_2}(s_{12}) \mathcal{R}_{b_2a_3a_4}
\bigg]\nonumber
\\
\label{eq.A5}
&\mathcal{A}_{a_1a_2a_3a_4a_5} = \mathcal{R}_{a_1a_2a_3a_4a_5} + 
\\
&+\frac{1}{5!}\sum_{S_5}\bigg[10 \,
\mathcal{R}_{a_1a_2a_3b_1}\;\Delta^{b_1b_2}(s_{45})\;\mathcal{R}_{b_2a_4a_5} 
\nonumber\\
&+15\;
\mathcal{R}_{a_1a_2b_1}\;\Delta^{b_1b_2}(s_{12})\;\mathcal{R}_{a_3b_2b_3}\;\Delta^{b_3b_4}(s_{45})\;\mathcal{R}_{b_4a_4a_5} 
\bigg]
\nonumber
\end{align}
The corresponding diagrammatic representations are provided, together with the result for $n=6$, in \S\ref{sec.amplitudes_results}. 
Besides combinatorial complexity, there is no conceptual obstacle to extending these formulas to $n\geq 7$. 

Our results for $n=3,\, 4$ agree with the literature (see \eg~\cite{Nagai:2019tgi,Cohen:2021ucp,Helset:2022tlf}) up to the geometrization of the scalar potential. The results for $n\leq  6$ agree with those in~\cite{Cheung:2021yog} in the  limit $V\to 0$ where the two can be compared. To our knowledge, the complete scalar potential contributions for $n\geq 5$ have not been reported before (although 
formally similar structures appear in 5-point amplitudes for scalar-gauge theories \cite{Helset:2022tlf}), which highlights the computational gain provided by our methods.

\subsubsection*{Cancellation of non-tensor terms}
It can be instructive to examine the cancellation of non-tensorial terms for $2\to 2$ scattering. The contribution of the non-tensor pieces to $\mathcal{A}_{a_1a_2a_3a_4}$ can be obtained replacing $\mathcal{R}\to F$ in Eq.~\eqref{eq.A4}, expanding the Feynman rules via~\eqref{eq.FvsR} and setting the four external legs on-shell. The last step makes the $N^{(2)}_{a_1a_2a_3a_4}$ terms vanish, leaving:
\begin{align}
\label{eq.A4_NT}
&\mathcal{A}_{a_1a_2a_3a_4}^{NT} \,\,=\,\, \bar N_{a_1a_2a_3a_4} 
\\
&+\frac{3}{4!}\sum_{S_4}\bigg[
N^{(2)}_{a_1a_2b_1} \Delta^{b_1b_2}(s_{12}) \mathcal{R}_{b_2a_3a_4}
\nonumber\\
&
+\mathcal{R}_{a_1a_2b_1} \Delta^{b_1b_2}(s_{12}) N^{(2)}_{b_2a_3a_4}
+N^{(2)}_{a_1a_2b_1} \Delta^{b_1b_2}(s_{12}) N^{(2)}_{b_2a_3a_4}
\bigg]\nonumber
\end{align}
We used that $\bar N_{a_1a_2a_3} =0$. This can be evaluated using the expressions for the non-tensor components of the 3- and 4-point Feynman rules in the fibre bundle picture:
\begin{align}
\label{eq.N2_3}
N^{(2)}_{a_1a_2a_3}&=
-\sum_{S_3}
\tensor{\Gamma}{^{b_1} _{a_1} _{a_2}}\Delta^{-1}_{b_1a_3}(p_3^2)  
\\
\label{eq.Nbar_4}
\bar N_{a_1a_2a_3a_4} &= \frac{3i}{4!}\sum_{S_4}\bigg[
\frac{\Lambda^4}{2}\tensor{R}{^\mu _{b_1} _\mu _{b_2}} \Gamma^{b_1}_{a_1 a_2} \Gamma^{b_2}_{a_3 a_4}
\\
&
+ s_{12}\Gamma_{b_1 a_1 a_2}\Gamma^{b_1}_{a_3 a_4} 
+ \Lambda^4\Gamma^{b_1}_{a_1 a_2} \nabla_{a_3} \tensor{R}{^\mu _{b_1} _\mu _{a_4}} \bigg]\nonumber
\end{align}
It is easy to check that, inserting these into~\eqref{eq.A4_NT}, together with~\eqref{eq.prop_FR} and~\eqref{eq.R3}, yields $\mathcal{A}_4^{NT}=0$. 
In particular, the contribution from the first line of Eq.~\eqref{eq.Nbar_4} cancels against the two contributions of the form $N\Delta \mathcal{R}$ in~\eqref{eq.A4_NT}, while the second line cancels the contribution $\sim N\Delta N$.

\subsubsection*{From trees to loops}

Finally, the covariant Feynman rules can also be used to obtain the loop corrections to lower-point vertices. To illustrate this, at 1-point and 2-point we have 1-loop corrections coming from the $\mathcal{R}_3$ and $\mathcal{R}_4$ vertices, which are:
\begin{align}
\mathcal{A}_{a_1}^{(1)} &= \int\frac{d^d k}{(2\pi)^d} \mathcal{R}_{a_1b_1b_2}\Delta^{b_1b_2}(k^2) 
\\
\mathcal{A}_{a_1a_2}^{(1)}&=
\frac{1}{2}\sum_{S_2}\int\frac{d^dk}{(2\pi)^d}\;\bigg[\mathcal{R}_{a_1a_2b_1b_1}\Delta^{b_1b_2}(k^2)
\\
&+\mathcal{R}_{a_1a_2b_1}\Delta^{b_1b_2}(0)\mathcal{R}_{b_2b_3b_4}\Delta^{b_3b_4}(k^2)
\nonumber\\
&+\mathcal{R}_{a_1b_1b_2}\mathcal{R}_{a_2b_3b_4}
\Delta^{b_1b_3}(k^2)\Delta^{b_2b_4}((p_1+k)^2)
\bigg]\, ,
\nonumber
\end{align}
matching the results of~\cite{Finn:2019aip}.
Extending this to higher points is beyond the scope of this paper, and will be treated in future work.

\section{Conclusions and Outlook}\label{sec.conclusions}

On-shell amplitudes in general scalar theories are built from quantities that are covariant under field redefinitions. This Letter demonstrated that, for tree-level calculations, one can obtain the amplitudes by adopting covariant Feynman rules from the outset and substituting these into a closed formula valid for any $n$-point amplitude.
For Lagrangians with generic 0- and 2-derivative interactions we show how these covariant Feynman rules can be efficiently derived from a metric on a fibre bundle, to high order in $n$.

Major computational obstacles due to the presence of non-tensorial terms in canonical Feynman rules are removed. This enables the calculation of amplitudes with many external legs, which we compute up to $n=10$. The resulting expressions are universal, and can be specialized to a given scalar theory by computing the relevant covariants as functions of its Lagrangian parameters (see the Supplemental Material). 
For fixed external states, the covariant Feynman rules are individually invariant, and so can themselves be interpreted as observable quantities.

These results constitute an important stepping stone towards the geometric interpretation of general EFTs. 
The next steps forward, that we will pursue in future work, include the extension of our formalism to loop amplitudes, including fermion and gauge fields, and the systematic inclusion of higher-derivative interactions by going from the fibre bundle to its $r$-jet bundle $J^r(E)$.
In that case, the entire EFT Lagrangian with up to $2(r+1)$ derivatives is obtained by pulling back a metric on $J^r(E)$~\cite{Alminawi:2023qtf}, and so the techniques developed here for efficiently building amplitudes starting from a single metric tensor should for the most part carry over. 

\medskip

\begin{acknowledgments}

We thank 
T. Cohen, A. Helset, Z. Polonsky and D. Sutherland
for discussions, and thank A.Helset and X.-X. Li for comments on the manuscript.
MA and IB acknowledge funding from SNF through the PRIMA grant no. 201508. 
The authors acknowledge support from the COMETA COST Action CA22130.
We are grateful to T. Cohen, X.-X. Li, and Z. Zhang for coordinating on~\cite{Cohen:2025toappear}.

\end{acknowledgments}

\bibliographystyle{apsrev4-1}
\bibliography{arxiv/bibliography}

\clearpage
\widetext
\appendix
\section{Covariant Feynman rules from fibre bundle geometry}
\label{sec.cFR_derivation}

The ``canonical" Feynman rules $F_{a_1\dots a_n}$ are $n$-point 1PI correlation functions derived from the tree-level effective action. In the fibre bundle formalism, this yields
\begin{align} \label{eq:can_feyn}
    &F_{a_1\dots a_n} =
    i\sum_{S_n} \left[\frac{\Lambda^4}{2} \eta^{\mu\nu}\;\de_{a_1}\dots \de_{a_n} g_{\mu\nu}  - \frac{1}{2} p_1\cdot p_2\;\de_{a_3}\dots \de_{a_n} g_{a_1a_2}\right]_{\phi_{\rm cl}}\, .
\end{align}
Covariant Feynman rules $\mathcal{R}_{a_1\dots a_n}$ can be obtained from here by expanding the metric and retaining only the covariant piece. The symmetries of the Riemann tensor ensure that the two blocks of the metric do not mix. In particular expressions such as $R^\mu_{a_i \nu a_j} R_{a_k a_l a_r a_s}$ vanish upon symmetrizing the expression in all its indices. Additionally note that $p_i^2 R_{a_i a_j a_k a_l}$ also vanishes after summing over permutations, due to the Bianchi identity.
Terms proportional to $\Gamma$ can safely be dropped as $\Gamma^\mu_{a_i \nu}|_{\text{vac}} =0$ and there are no 3-index tensors when using the Levi-Civita connection, so $\Gamma_{a_i a_j a_k}$ terms will always cancel. In the derivation of covariant rules we implement substitutions that are equal up to powers of $\Gamma$.

To derive the $g_{\mu \nu}$ contributions to the covariant Feynman rules we use a number of simplifying tricks. First, we exchange derivatives of the metric for Christoffel symbols; then, using also Poincar\'e invariance, we find
\begin{align}
     &\tensor{R}{^\mu _{a_i} _\nu _{a_j}} = -\partial_{a_j} \Gamma^\mu_{a_i \nu} + \Gamma^\mu_{\nu a_m} \Gamma^{a_m}_{a_i a_j} -  \Gamma^\mu_{a_j \rho} \Gamma^\rho_{a_i\nu}
    \label{mixed riemann tensor definition}
\end{align}
which allows us to define higher derivatives of $\Gamma^\mu_{a_i \nu}$ recursively. Since the blocks of the metric do not mix and only vertical tensors contribute to the amplitude we find $\partial^n \tensor{R}{^\mu _{a_i} _\nu _{a_j}} \to \nabla^n\tensor{R}{^\mu _{a_i} _\nu _{a_j}}$ for any $n$.

Now we turn our attention to the contributions from $g_{a_i a_j}$ to the covariant Feynman rules. The symmetries of the Riemann tensor make the theory behave as though it were massless, in that we can replace $p_i \cdot p_j \to \frac{1}{2}s_{ij}$. Momentum conservation, 
$\sum_{i \neq j} s_{ij}=0$,
allows us to eliminate $n$ variables from a set of $\binom{n}{2}$. It is convenient to eliminate the variables $s_{in}$ for $1 \leq i \leq n-1$ as well as the variable $s_{n-1 n-2}$, allowing the substitution
\begin{align}
    &R_{a_1 a_3 a_4 a_2} + \Gamma_{q a_1 a_4}\Gamma^q_{a_2 a_3} - \Gamma_{q a_1 a_2}\Gamma^{q}_{a_3 a_4} \\&\to \frac{1}{2}(g_{a_1 a_2,a_3 a_4} + g_{a_3 a_4, a_1 a_2} - g_{a_2 a_3,a_1 a_4} - g_{a_1 a_4, a_2 a_3}), 
\end{align}
which, in addition to using the Bianchi identity, means the contribution to~\eqref{eq:can_feyn} from the kinetic term reads
\begin{align}
    & \sum_{S_n}\frac{1}{n!} \bigg(\binom{n}{2}- n \bigg)s_{12}\partial_{a_3} \dots \partial_{a_{n-2}} 
    \bigg(R_{a_1 a_{n-1} a_n a_2} + \Gamma_{q a_1 a_n}\Gamma^q_{a_2 a_{n-1}} - \Gamma_{q a_1 a_2}\Gamma^{q}_{a_n a_{n-1}}\bigg)\, . 
    \label{n-point kinetic term differential equation}
\end{align}
To go to $n \geq 5$ we need to then systematically replace partial derivatives of the Riemann tensor and Christoffel symbols with tensors. To do so we observe that, somewhat intriguingly, momentum conservation imposes similar conditions on the Christoffel symbols as switching to geodesic coordinates would do. For example, of relevance to $n=4$ points, we have
$\partial_{a_4}\Gamma_{a_1 a_2 a_3}(s_{12} + s_{13} + s_{14}) \rightarrow 0 \nonumber$. By permuting indices, this implies
$s_{12}(\partial_{a_4}\Gamma_{a_1 a_2 a_3} + \partial_{a_3}\Gamma_{a_1 a_2 a_4} +\partial_{a_2}\Gamma_{a_1 a_4 a_3}) \rightarrow 0$, enabling us to substitute $s_{12} \partial_{a_2} \Gamma_{a_1 a_3 a_4} \rightarrow -\frac{1}{3}s_{12}(R_{a_1 a_3 a_4 a_2} + R_{a_1 a_4 a_3 a_2})$.

This kind of replacement rule can be generalised to arbitrary $n$, where it reads
\begin{align}
   &\sum_{S_{n-2}} s_{12}\binom{n-2}{2} \partial_{a_n} \dots \partial_{a_2} \Gamma_{a_1 a_3 a_4} %\nonumber \\ & 
   \rightarrow - \sum_{S_{n-2}}s_{12} \binom{n-2}{1} \partial_{a_n} \dots \partial_{a_4} \Gamma_{a_1 a_2 a_3}\, . 
\end{align}
Finally, we caution that when going beyond $n \geq 6$ it becomes important to distinguish between Christoffel symbols with the first index raised or lowered:
\begin{align}
    &s_{12} \partial_{a_6} \partial_{a_5} \partial_{a_2} \Gamma_{a_1 a_3 a_4} 
    % \\&
    \to -s_{12}(\frac{3}{5}\nabla_{a_6} \nabla_{a_5}R_{a_1 a_3 a_4 a_2} + \frac{8}{15}R_{q a_5 a_6 a_1}\tensor{R}{^q _{a_3} _{a_4} _{a_2}}), 
\end{align}
whereas
\begin{align}
    &s_{12} \partial_{a_6} \partial_{a_5} \partial_{a_2} \Gamma^{a_1}_{a_3 a_4} 
    % \\&
    \to -s_{12}(\frac{3}{5}\nabla_{a_6} \nabla_{a_5}\tensor{R}{^{a_1} _{a_3} _{a_4} _{a_2}} - \frac{2}{15}  \tensor{R}{^{a_1} _{a_3} _{a_4} _{q}}\tensor{R}{^{q} _{a_5} _{a_6} _{a_2}})\, . 
\end{align}
As advertised, these conditions are identical to those we get from going to Riemann normal coordinates, and formally lead to the same metric expansion as verified by comparing our results with \cite{Muller:1999rnc} up to $n=10$. However, it is important to reiterate that these conditions in our case originate from imposing the physical condition of momentum conservation, and that regardless of the coordinate system of choice the covariant Feynman rules are not the same as the ``canonical" Feynman rules of the theory. However, they both yield the same results for the S-matrix elements.

The results for $6\leq n\leq 10$ are:

\allowdisplaybreaks
\begin{align}
\label{eq.R6}
&\mathcal{R}_{a_1\dots a_6} = \frac{i}{6!}\sum_{S_6}\bigg[
\frac{\Lambda^4}{2}\nabla_{a_1}\nabla_{a_2}\nabla_{a_3}\nabla_{a_4}\tensor{R}{^\mu_{a_5}_\mu_{a_6}}
-11 \Lambda^4\tensor{R}{^\mu_{a_1}_\nu_{a_2}} \nabla_{a_3}\nabla_{a_4}\tensor{R}{^\nu_{a_5}_\mu_{a_6}}
- 7 \Lambda^4\nabla_{a_1}\tensor{R}{^\mu_{a_2}_\nu_{a_3}}\nabla_{a_4}\tensor{R}{^\nu_{a_5}_\mu_{a_6}}
\nonumber\\
&
+ 8 \Lambda^4\tensor{R}{^\mu_{a_1}_\nu_{a_2}}\tensor{R}{^\nu_{a_3}_\rho_{a_4}}\tensor{R}{^\rho_{a_5}_\mu_{a_6}}
-9 s_{12} \, \nabla_{a_5}\nabla_{a_6}R_{a_1a_4a_3a_2}
-8 s_{12} \, \tensor{R}{_{a_1}_{a_3}_{a_4}^{b_1}}\tensor{R}{_{b_1}_{a_5}_{a_6}_{a_2}}
\bigg]\, ,
\\
\label{eq.R7}
&\mathcal{R}_{a_1\dots a_7} = 
\frac{i}{7!}\sum_{S_7}\bigg[
\frac{\Lambda^4}{2}\nabla_{a_1}\nabla_{a_2}\nabla_{a_3}\nabla_{a_4}\nabla_{a_5}\tensor{R}{^\mu_{a_6}_\mu_{a_7}}
-16 \Lambda^4\tensor{R}{^\mu_{a_1}_\nu_{a_2}} \nabla_{a_3}\nabla_{a_4}\nabla_{a_5}\tensor{R}{^\nu_{a_6}_\mu_{a_7}}
-25 \Lambda^4\nabla_{a_1}\tensor{R}{^\mu_{a_2}_\nu_{a_3}} \nabla_{a_4}\nabla_{a_5}\tensor{R}{^\nu_{a_6}_\mu_{a_7}}
\nonumber\\
&
+ 64 \Lambda^4\tensor{R}{^\mu_{a_1}_\nu_{a_2}}\tensor{R}{^\nu_{a_3}_\rho_{a_4}}\nabla_{a_5}\tensor{R}{^\rho_{a_6}_\mu_{a_7}}
-14 s_{12} \, \nabla_{a_5}\nabla_{a_6}\nabla_{a_7} R_{a_1a_4a_3a_2}
-56 s_{12} \, \tensor{R}{_{a_1}_{a_3}_{a_4}^{b_1}}\nabla_{a_5}\tensor{R}{_{b_1}_{a_6}_{a_7}_{a_2}}
\bigg]\, ,
\\
\label{eq.R8}
&\mathcal{R}_{a_1\dots a_8} = 
\frac{i}{8!}\sum_{S_8}\bigg[
\frac{\Lambda^4}{2}\nabla_{a_1}\nabla_{a_2}\nabla_{a_3}\nabla_{a_4}\nabla_{a_5}\nabla_{a_6}\tensor{R}{^\mu_{a_7}_\mu_{a_8}}
-22 \Lambda^4 \tensor{R}{^\mu_{a_1}_\nu_{a_2}} \nabla_{a_3}\nabla_{a_4}\nabla_{a_5}\nabla_{a_6}\tensor{R}{^\nu_{a_7}_\mu_{a_8}}
\nonumber\\
&
-41\Lambda^4 \nabla_{a_1}\tensor{R}{^\mu_{a_2}_\nu_{a_3}}\nabla_{a_4}\nabla_{a_5}\nabla_{a_6}\tensor{R}{^\nu_{a_7}_\mu_{a_8}}
-25\Lambda^4 \nabla_{a_1}\nabla_{a_2}\tensor{R}{^\mu_{a_3}_\nu_{a_4}}\nabla_{a_5}\nabla_{a_6}\tensor{R}{^\nu_{a_7}_\mu_{a_8}}
+ 144 \Lambda^4 \tensor{R}{^\mu_{a_1}_\nu_{a_2}}\tensor{R}{^\nu_{a_3}_\rho_{a_4}}\nabla_{a_5}\nabla_{a_6}\tensor{R}{^\rho_{a_7}_\mu_{a_8}}
\nonumber\\
&
+ 182 \Lambda^4 \tensor{R}{^\mu_{a_1}_\nu_{a_2}}\nabla_{a_3}\tensor{R}{^\nu_{a_4}_\rho_{a_5}}\nabla_{a_6}\tensor{R}{^\rho_{a_7}_\mu_{a_8}} - 32 \tensor{R}{^\mu_{a_1}_\nu_{a_2}} \tensor{R}{^\nu_{a_3}_\rho_{a_4}} \tensor{R}{^\mu_{a_5}_\sigma_{a_6}} \tensor{R}{^\sigma_{a_7}_\rho_{a_8}}
- 20 s_{12} \nabla_{a_3}\nabla_{a_4}\nabla_{a_5}\nabla_{a_6}\tensor{R}{_{a_1}_{a_7}_{a_8}_{a_2}}
\nonumber\\
&
- 136 s_{12} 
\tensor{R}{_{a_1}_{a_3}_{a_4}_{b_1}}\nabla_{a_5}\nabla_{a_6}\tensor{R}{^{b_1}_{a_7}_{a_8}_{a_2}}- 110 s_{12} 
\nabla_{a_3}\tensor{R}{_{a_1}_{a_4}_{a_5}^{b_1}}\nabla_{a_6}\tensor{R}{_{b_1}_{a_7}_{a_8}_{a_2}} - 32s_{12} \tensor{R}{_{a_1}_{a_3}_{a_4}^{b_1}} \tensor{R}{_{b_1}_{a_5}_{a_6}^{b_2}} \tensor{R}{_{a_2}_{a_7}_{a_8}_{b_2}}
\bigg],
\\
\label{eq.R9}
&\mathcal{R}_{a_1\dots a_9} = 
\frac{i}{9!}\sum_{S_9}\bigg[
\frac{\Lambda^4}{2}\nabla_{a_1}\nabla_{a_2}\nabla_{a_3}\nabla_{a_4}\nabla_{a_5}\nabla_{a_6}\nabla_{a_7}\tensor{R}{^\mu_{a_8}_\mu_{a_9}}
-29 \Lambda^4 \tensor{R}{^\mu_{a_1}_\nu_{a_2}} \nabla_{a_3}\nabla_{a_4}\nabla_{a_5}\nabla_{a_6}\nabla_{a_7}\tensor{R}{^\nu_{a_8}_\mu_{a_9}}
\nonumber\\
&
-63\Lambda^4 \nabla_{a_1}\tensor{R}{^\mu_{a_2}_\nu_{a_3}}\nabla_{a_4}\nabla_{a_5}\nabla_{a_6}\nabla_{a_7}\tensor{R}{^\nu_{a_8}_\mu_{a_9}}
-91\Lambda^4 \nabla_{a_1}\nabla_{a_2}\tensor{R}{^\mu_{a_3}_\nu_{a_4}}\nabla_{a_5}\nabla_{a_6}\nabla_{a_7}\tensor{R}{^\nu_{a_8}_\mu_{a_9}}
\\&+ 284 \Lambda^4 \tensor{R}{^\mu_{a_1}_\nu_{a_2}}\tensor{R}{^\nu_{a_3}_\rho_{a_4}}\nabla_{a_5}\nabla_{a_6}\nabla_{a_7}\tensor{R}{^\rho_{a_8}_\mu_{a_9}}
+ 890 \Lambda^4 \tensor{R}{^\mu_{a_1}_\nu_{a_2}}\nabla_{a_3}\tensor{R}{^\nu_{a_4}_\rho_{a_5}}\nabla_{a_6}\nabla_{a_7}\tensor{R}{^\rho_{a_8}_\mu_{a_9}}
\nonumber\\
&+ 192 \Lambda^4 \nabla_{a_1}\tensor{R}{^\mu_{a_2}_\nu_{a_3}}\nabla_{a_4}\tensor{R}{^\nu_{a_5}_\rho_{a_6}}\nabla_{a_7}\tensor{R}{^\rho_{a_8}_\mu_{a_9}}
- 464 \tensor{R}{^\mu_{a_1}_\nu_{a_2}} \tensor{R}{^\nu_{a_3}_\rho_{a_4}} \tensor{R}{^\mu_{a_5}_\sigma_{a_6}} \nabla_{a_7}\tensor{R}{^\sigma_{a_8}_\rho_{a_9}} - 27 s_{12} \nabla_{a_3}\nabla_{a_4}\nabla_{a_5}\nabla_{a_6}\nabla_{a_7}\tensor{R}{_{a_1}_{a_8}_{a_9}_{a_2}}
\nonumber \\
&- 276 s_{12} 
\tensor{R}{_{a_1}_{a_3}_{a_4}^{b_1}}\nabla_{a_5}\nabla_{a_6}\nabla_{a_7}\tensor{R}{_{b_1}_{a_8}_{a_9}_{a_2}}- 594 s_{12} 
\nabla_{a_3}\tensor{R}{_{a_1}_{a_4}_{a_5}^{b_1}}\nabla_{a_6}\nabla_{a_7}\tensor{R}{_{b_1}_{a_8}_{a_9}_{a_2}}  - 432 s_{12} \tensor{R}{_{b_1}_{a_3}_{a_4}_{a_1}} \tensor{R}{^{b_1}_{a_5}_{a_6}_{b_2}} \nabla_{a_7}\tensor{R}{^{b_2}_{a_8}_{a_9}_{a_2}}
\bigg]\, ,\nonumber\\
\label{eq.R10}
&\mathcal{R}_{a_1\dots a_{10}} = 
\frac{i}{10!}\sum_{S_{10}}\bigg[
\frac{\Lambda^4}{2}\nabla_{a_1}\nabla_{a_2}\nabla_{a_3}\nabla_{a_4}\nabla_{a_5}\nabla_{a_6}\nabla_{a_7}\nabla_{a_8}\tensor{R}{^\mu_{a_9}_\mu_{a_{10}}}
-37 \Lambda^4 \tensor{R}{^\mu_{a_1}_\nu_{a_2}} \nabla_{a_3}\nabla_{a_4}\nabla_{a_5}\nabla_{a_6}\nabla_{a_7}\nabla_{a_8}\tensor{R}{^\nu_{a_9}_\mu_{a_{10}}}
\nonumber\\
&
-92\Lambda^4 \nabla_{a_1}\tensor{R}{^\mu_{a_2}_\nu_{a_3}}\nabla_{a_4}\nabla_{a_5}\nabla_{a_6}\nabla_{a_7}\nabla_{a_8}\tensor{R}{^\nu_{a_9}_\mu_{a_{10}}}
-154\Lambda^4 \nabla_{a_1}\nabla_{a_2}\tensor{R}{^\mu_{a_3}_\nu_{a_4}}\nabla_{a_5}\nabla_{a_6}\nabla_{a_7}\nabla_{a_8}\tensor{R}{^\nu_{a_9}_\mu_{a_{10}}}\nonumber \\
&
-91\Lambda^4 \nabla_{a_1}\nabla_{a_2}\nabla_{a_3}\tensor{R}{^\mu_{a_4}_\nu_{a_5}}\nabla_{a_6}\nabla_{a_7}\nabla_{a_8}\tensor{R}{^\mu_{a_9}_\nu_{a_{10}}}
+508 \Lambda^4 \tensor{R}{^\mu_{a_1}_\nu_{a_2}}\tensor{R}{^\nu_{a_3}_\rho_{a_4}}\nabla_{a_5}\nabla_{a_6}\nabla_{a_7}\nabla_{a_8}\tensor{R}{^\rho_{a_9}_\mu_{a_{10}}}
\nonumber\\
&
+1918 \Lambda^4 \tensor{R}{^\mu_{a_1}_\nu_{a_2}}\nabla_{a_3}\tensor{R}{^\nu_{a_4}_\rho_{a_5}}\nabla_{a_6}\nabla_{a_7}\nabla_{a_8}\tensor{R}{^\rho_{a_9}_\mu_{a_{10}}}
+1436 \Lambda^4 \nabla_{a_1}\tensor{R}{^\mu_{a_2}_\nu_{a_3}}\nabla_{a_4}\tensor{R}{^\nu_{a_5}_\rho_{a_6}}\nabla_{a_7}\nabla_{a_8}\tensor{R}{^\rho_{a_9}_\mu_{a_{10}}}
\nonumber \\
&
+1178 \Lambda^4 \tensor{R}{^\mu_{a_1}_\nu_{a_2}}\nabla_{a_3}\nabla_{a_4}\tensor{R}{^\nu_{a_5}_\rho_{a_6}}\nabla_{a_7}\nabla_{a_8}\tensor{R}{^\rho_{a_9}_\mu_{a_{10}}}
-1360 \Lambda^4 \tensor{R}{^\mu_{a_1}_\nu_{a_2}}\tensor{R}{^\nu_{a_3}_\rho_{a_4}}\tensor{R}{^\rho_{a_5}_\sigma_{a_6}}\nabla_{a_7}\nabla_{a_8}\tensor{R}{^\sigma_{a_9}_\mu_{a_{10}}}
\nonumber \\
&
-2648 \Lambda^4 \tensor{R}{^\mu_{a_1}_\nu_{a_2}}\tensor{R}{^\nu_{a_3}_\rho_{a_4}}\nabla_{a_5}\tensor{R}{^\rho_{a_6}_\sigma_{a_7}}\nabla_{a_8}\tensor{R}{^\sigma_{a_9}_\mu_{a_{10}}} + 128 \Lambda^4 \tensor{R}{^\mu_{a_1}_\nu_{a_2}}\tensor{R}{^\nu_{a_3}_\rho_{a_4}}\tensor{R}{^\rho_{a_5}_\sigma_{a_6}}\tensor{R}{^\sigma_{a_7}_\alpha_{a_8}}\tensor{R}{^\alpha_{a_9}_\mu_{a_{10}}}
\nonumber\\
&
- 35 s_{12} \nabla_{a_3}\nabla_{a_4}\nabla_{a_5}\nabla_{a_6}\nabla_{a_7} \nabla_{a_8}\tensor{R}{_{a_1}_{a_9}_{a_{10}}_{a_2}}
- 500 s_{12} 
\tensor{R}{_{a_1}_{a_3}_{a_4}_{b_1}}\nabla_{a_5}\nabla_{a_6}\nabla_{a_7} \nabla_{a_8}\tensor{R}{^{b_1}_{a_9}_{a_{10}} _{a_2}}
\nonumber \\
&- 1330 s_{12} 
\nabla_{a_3}\tensor{R}{_{a_1}_{a_4}_{a_5} _{b_1}}\nabla_{a_6}\nabla_{a_7} \nabla_{a_8}\tensor{R}{^{b_1}_{a_9}_{a_{10}}_{a_2}} - 882 s_{12} \nabla_{a_3} \nabla_{a_4} \tensor{R}{_{a_1}_{a_5}_{a_6} _{b_1}}\nabla_{a_7} \nabla_{a_8}\tensor{R}{^{b_1}_{a_9}_{a_{10}}_{a_2}} \nonumber \\
&- 1328 s_{12}  \tensor{R}{_{a_1}_{a_3}_{a_4} _{b_1}} \tensor{R}{^{b_1}_{a_5}_{a_6} _{b_2}}\nabla_{a_7} \nabla_{a_8}\tensor{R}{^{b_2}_{a_9}_{a_{10}}_{a_2}} - 2120 s_{12}  \tensor{R}{_{a_1}_{a_3}_{a_4} _{b_1}} \nabla_{a_5}\tensor{R}{^{b_1}_{a_6}_{a_7} _{b_2}}\nabla_{a_8} \tensor{R}{^{b_2}_{a_9}_{a_{10}}_{a_2}} \nonumber\\
& - 128 s_{12}  \tensor{R}{_{a_1}_{a_3}_{a_4} _{b_1}}   \tensor{R}{^{b_1}_{a_5}_{a_6} _{b_2}}   \tensor{R}{^{b_2}_{a_7}_{a_8} _{b_3}}   \tensor{R}{^{b_3}_{a_9}_{a_{10}} _{a_2}} 
\bigg]\, .
\end{align}

\newpage
\section{Tree-level on-shell amplitude with $n=6$}\label{sec.amplitudes_results}
\begin{figure}[t]\centering
\includegraphics[width=8cm]{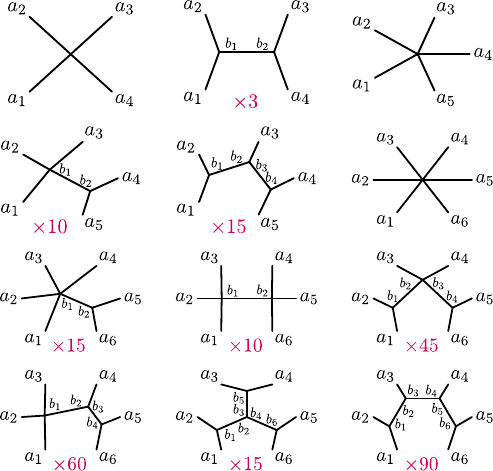}
\caption{Diagrams for the construction of on-shell amplitudes $\mathcal{A}_n$ with $n=4,5,6$ via covariant Feynman rules. The number of independent index contractions for each topology is shown in magenta.}\label{fig.A4_A6}
\end{figure}

The expression for the $n=6$ on-shell amplitude as a function of covariant Feynman rules is:
\begin{align}
\mathcal{A}_{a_1a_2a_3a_4a_5a_6} &= \mathcal{R}_{a_1a_2a_3a_4a_5a_6} 
\\
&
+\frac{1}{6!}\sum_{S_6}\bigg[
15\,
\mathcal{R}_{a_1a_2a_3a_4b_1}\,\Delta^{b_1b_2}(s_{56})\,\mathcal{R}_{b_2a_5a_6}  
\nonumber\\
&+
10\,
\mathcal{R}_{a_1a_2a_3b_1}\,\Delta^{b_1b_2}(s_{123})\,\mathcal{R}_{b_2a_4a_5a_6} 
\nonumber\\
&+
45\,
\mathcal{R}_{a_1a_2b_1}\Delta^{b_1b_2}(s_{12})\,
\mathcal{R}_{a_3a_4b_2b_3}\,\,\Delta^{b_3b_4}(s_{56})\,\mathcal{R}_{b_4a_5a_6} 
\nonumber\\
&
+
60\,
\mathcal{R}_{a_1a_2a_3b_1}\,\Delta^{b_1b_2}(s_{1234})\,\mathcal{R}_{b_2b_3a_4}\,\Delta^{b_3b_4}(s_{56})\,\mathcal{R}_{b_4a_5a_6} 
\nonumber\\
&
+
15\,
\mathcal{R}_{a_1a_2b_1}\,\Delta^{b_1b_2}(s_{12})\,\mathcal{R}_{b_2b_3b_4}\,\Delta^{b_3b_5}(s_{34})\,\mathcal{R}_{b_5a_3a_4}\,
\Delta^{b_4b_6}(s_{56})\,
\mathcal{R}_{b_6a_5a_6} \nonumber
\\
&+ 
90\,
\mathcal{R}_{a_1a_2b_1}\,\Delta^{b_1b_2}(s_{12})\,\mathcal{R}_{b_2b_3a_3}\,\Delta^{b_3b_4}(s_{123})\,\mathcal{R}_{b_4b_5a_4}\,
\Delta^{b_5b_6}(s_{56})\, \mathcal{R}_{b_6a_5a_6} \,\,\,
\bigg]\nonumber
\end{align}
Figure~\ref{fig.A4_A6} shows diagrammatic representations for amplitudes with $4\leq n\leq 6$.

\clearpage
\widetext 

\setcounter{equation}{0}
\renewcommand\theequation{S\arabic{equation}} 

\section*{Supplemental material}

\subsection*{Covariance in 0-d toys} 
\label{subsec: cov 0d}
Consider a zero-dimensional quantum field theory (see {\em e.g.}~\cite{skinner2018quantum}). This drastic assumption brings a number of simplifications -- not least rendering kinematics trivial, and reducing functional derivatives to partial derivatives. But it nevertheless preserves the key combinatorial properties of the Feynman graph expansion which guarantee the covariance of amplitudes, which we here illustrate with the 4-point amplitude. 

In zero dimensions the scalar field is a map $\phi:\text{pt} \to \cM$ {\em i.e.} just an $\cM$-valued variable, no longer a map (or section). 
This means that a field redefinition $\phi \mapsto \psi(\phi)$ is, in this 0-d setting, {\em exactly the same} as a diffeomorphism on field space $\cM$. So, amplitudes are precisely tensors on $\cM$. 
The action is just an ordinary function (not functional) of the field variable $\phi$, and correlation functions are just expectation values of operators on the probability distribution defined by $e^{-S(\phi)/\hbar}$, suitably normalised.
For instance, we could take
    $S(\phi) = \frac{1}{2} m^2 \phi^2 + \lambda\phi^4 + \dots$
assuming the highest power in $\phi$ to be even for convergence of path integrals.
The amputated 1PI $n$-point functions, together with the `propagator' (even though nothing propagates in zero-dimensions), are now just partial derivatives of the effective action $\Gamma = S + \dots$,
\begin{equation}
    \langle \phi_1 \dots \phi_n \rangle_{\text{1PI}} = \frac{i\partial^n \Gamma}{\partial \phi_1 \dots \partial \phi_n}\, , \qquad 
    D_{ij} = -\left(\frac{i\partial^2 \Gamma}{\partial \phi_i \partial \phi_j}\right)^{-1}\, .
\end{equation}
Under a general field redefinition $\phi \mapsto \psi(\phi)$, the variation of the 2-point function is given by the chain rule:
\begin{equation}
    \frac{\partial^2 \Gamma(\psi)}{\partial \phi \partial \phi} = \frac{\partial^2 \Gamma(\psi)}{\partial \psi \partial \psi} \frac{\partial \psi}{\partial \phi}\frac{\partial \psi}{\partial \phi} + \frac{\partial \Gamma(\psi)}{\partial \psi} \frac{\partial^2 \psi}{\partial \phi^2}\, ,
\end{equation}
where we drop indices in the remainder of this Section for readability.
Imposing the vacuum condition $\partial_\psi\Gamma=0$ kills the second term, meaning that the 2-point, and hence its inverse (the propagator), transforms as a tensor on $\cM$.

Consider the four-point amplitude. It is
\begin{align} \label{eq:0d-4pt}
    i\mathcal{A}_4\, =  \quad   
    \begin{tikzpicture}
    [baseline=-1.0ex]
    \begin{feynman}
    \vertex (a);
    \vertex [above left=0.2in of a] (i1);
    \vertex [below left=0.2in of a] (i2);
    \vertex [above right=0.2in of a] (j1);
    \vertex [below right=0.2in of a] (j2);
    \diagram* {
      (i1) -- (a), 
      (i2) -- (a),
      (a) -- (j1),
      (a) -- (j2),
    };
    \end{feynman}
    \end{tikzpicture}
    \quad + \quad
    \begin{tikzpicture}
    [baseline=-1.0ex]
    \begin{feynman}
    \vertex (a);
    \vertex [above left=0.2in of a] (i1);
    \vertex [below left=0.2in of a] (i2);
    \vertex [right=0.2in of a] (b);
    \vertex [above right=0.2in of b] (j1);
    \vertex [below right=0.2in of b] (j2);
    \diagram* {
      (i1) -- (a), 
      (i2) -- (a),
      (a) -- (b),
      (b) -- (j1),
      (b) -- (j2),
    };
    \end{feynman}
    \end{tikzpicture} 
    \quad 
    = i\Gamma^{(4)} - \Gamma^{(3)}\frac{i}{\Gamma^{(2)}} \Gamma^{(3)},
\end{align}
where we adopt the notation $f^{(n)}(x)\equiv \frac{\partial^n f(x)}{\partial x^n}$.
Note that there is no kinematics so, contrary to our usual diagrammatic thinking, the second diagram does not correspond to a particular `channel' and is unique.
Under $\phi \mapsto \psi(\phi)$, the 4-point contact has variation determined by the Fa\`a di Bruno formula~\eqref{eq:FdB}, which is just the result of successively applying the chain rule:
\begin{equation}
     \frac{\partial^4 \Gamma(\psi)}{\partial \phi \partial \phi\partial \phi \partial \phi}  = \Gamma^{(4)}(\psi) \left(\frac{\partial \psi}{\partial \phi}\right)^4 + \textcolor{blue}{6\Gamma^{(3)}(\psi)\left(\frac{\partial \psi}{\partial \phi}\right)^2 \frac{\partial^2 \psi}{\partial \phi^2} } + \textcolor{magenta}{\Gamma^{(2)}(\psi) \left(3\left(\frac{\partial^2 \psi}{\partial \phi^2}\right)^2   + 4 \frac{\partial^3 \psi}{\partial \phi^3} \frac{\partial \psi}{\partial \phi}\right)} + \textcolor{red}{\Gamma^{(1)}(\phi) \frac{\partial^4 \psi}{\partial \phi^4}}
\end{equation}
The last term in red vanishes due to the vacuum condition $\textcolor{red}{\Gamma^{(1)}=0}$. 
The penultimate term in magenta vanishes only if we impose the on-shell condition $\textcolor{magenta}{\Gamma^{(2)}=0}$. This leaves only the black and blue terms. The black is the transformation of a tensor, while the blue is not.
The leading variation of the second diagram in Eq.~\eqref{eq:0d-4pt}, from gluing lower points, precisely cancels the term in blue. Using the Fa\`a di Bruno on both 3-point insertions:
\begin{align}
    \left(\frac{\partial^2 \Gamma(\psi) }{\partial\phi\partial\phi}\right)^{-1} \left(\frac{\partial^3 \Gamma(\psi) }{\partial\phi\partial\phi\partial\phi} \right)^2
    &= \frac{1}{\Gamma^{(2)}(\phi)} \left(\frac{\partial \phi}{\partial \psi}\right)^2 \left[\Gamma^{(3)}(\psi) \left(\frac{\partial \psi}{\partial \phi}\right)^3 + 3 \Gamma^{(2)}(\psi)\frac{\partial^2 \psi}{\partial \phi^2}\frac{\partial \psi}{\partial \phi}  + \textcolor{red}{\Gamma^{(1)}(\phi) \frac{\partial^3 \psi}{\partial \phi^3}} \right]^2 \\
    &= \frac{(\Gamma^{(3)}(\psi))^2}{\Gamma^{(2)}(\psi)} \left(\frac{\partial \psi}{\partial \phi}\right)^4 + \textcolor{blue}{6\Gamma^{(3)}(\psi)\left(\frac{\partial \psi}{\partial \phi}\right)^2 \frac{\partial^2 \psi}{\partial \phi^2} } + \textcolor{magenta}{9\Gamma^{(2)}(\psi)\left(\frac{\partial^2 \psi}{\partial \phi^2}\right)^2} + \textcolor{red}{\Gamma^{(1)}(\psi)[\dots]}
\end{align}
So the glued diagram contains a tensor piece (black) plus a non-tensor piece (blue) that precisely matches that coming from the variation of the contact term. The pink and red non-tensor pieces do not cancel, 
and are only removed by the on-shell and vacuum conditions. 
The upcoming paper~\cite{Alminawi:future}
will demonstrate, restoring full $d$-dimensional kinematics, 
that similar cancellations occur for all non-tensor pieces at $n$-points, for all finite $n\in \mathbb{Z}_{\geq 2}$. In other words, the expansion of the $n$-point amplitude in Feynman diagrams coincides precisely with a diagrammatic representation of the Fa\`a di Bruno formula for the transformation of the $n^{\text{th}}$ derivative, a correspondence that only breaks down only for the terms linear in $\Gamma^{(1)}$ and $\Gamma^{(2)}$.

\subsection*{Vertical Tensors on Fibre Bundles}

Given our field space fibre bundle $(E,\pi,\Sigma)$, the quantum field theory amplitudes $\mathcal{A}_n$ are expansions in vertical tensors on that bundle. Here, for completeness, we recap how vertical tensors are constructed. We note that the approach taken in Ref.~\cite{Craig:2023hhp} for distinguishing fibre {\em vs.} spacetime directions is similar in spirit.

At any point $e\in E$, we can split the tangent space $T_e E$ into vertical and horizontal directions using the projection $\pi$ (see Fig.~\ref{fig:Ve}). The vertical space $V_e E$ is well-defined, spanned by the tangent vectors at $e$ that get mapped to zero under $\pi$, $V_e E := \ker (d\pi_e)$,
where $d\pi_e:T_e E \to T_{\pi(e)}\Sigma$ is the differential of the map $\pi$ at $e$, which recall is a linear approximation of $\pi$ {\em i.e.} provides a map $d\pi_e:T_e E \to T_{\pi(e)}\Sigma$ between tangent spaces. 
Collecting the $\{V_e E\}$ as fibres over $\Sigma$, we form the vertical bundle $VE=\ker(d\pi)\subset TE$, with projection map $\pi_V:VE\to \Sigma$.
From here, we can tensor product (and/or dualise) vertical vectors to build {\em vertical tensors} $T_V$.

\medskip

In order to correctly reproduce the scalar amplitudes using our fibre bundle formalism, we need to build vertical tensors that are Lorentz invariant but which nevertheless end up depending on the $g_{\mu\nu}$ components of the metric, such as 
that in~\eqref{eq:Rij}. To form such tensors requires not just the vertical bundle $VE$ but also a horizontal bundle $HE$. While $VE$ is uniquely defined given $\pi$, specifying $HE$ amounts to a {\em choice} of Ehresmann connection on the bundle, {\em i.e.} a complementary `horizontal' subspace $H_e E$ for every $e\in E$ such that 
$T_e E = V_e E \oplus H_e E$ (see Fig.~\ref{fig:Ve}),
and thence a global splitting
$TE=VE\oplus HE$. See Fig.~\ref{fig:Ve}.

With this splitting, one can then resolve tangent vectors on $E$ into vertical and horizontal directions. 
Using the connection we can also {\em lift} horizontal vector fields from the base $\Sigma$ to $HE$. Starting from an orthonormal basis $\{e_\mu\}$ of vector fields on $\Sigma$, we let $\{\tilde e_\mu^H\}$ denote their lifts to $HE$.
We make the convenient choice of Ehresmann connection defined in local coordinates via $\tilde e_\mu^H = \partial_\mu$. We can also take the horizontal lift of a covector $e^\mu:=g^{\mu I} e_J$, which we denote $\tilde e^{\mu H}$. Note the covector is first defined by raising with the full bundle metric $g$.

Equipped with these objects, we can explain how a vertical tensor like~\eqref{eq:Rij} in the main text is formally defined.
To form $\tilde R$, we start from the Riemann tensor $R$ on $E$, which is of type $(1,3)$, and define $\tilde R$ by specifying the following action on a pair of vertical vector fields, 
    $\tilde R (v_1,v_2) := \pi_V R(\tilde{e}^{\mu H}, v_1, \tilde e_\mu^H, v_2)$,
where $v_{1,2}\in \Gamma(VE)$. The result is
    $\tilde R_{ij} = \tensor{R}{^\mu _i _\mu _j} = \frac{1}{2\Lambda^4}\mathrm{Tr}[\eta]\partial_{i}\partial_j V$
as quoted in the main text.

\subsection*{Explicit covariant Feynman rules in a theory of two real scalars}

To better appreciate the physical meaning of the results presented in the main text, it can be useful to evaluate the covariant expressions for a concrete theory.
To this end, we consider a theory of two real scalar fields $\phi_1, \phi_2$. Following Sec.~6.3 of Ref.~\cite{Alminawi:2023qtf}, we parameterize the most general fibre bundle metric through the following Taylor expansions:
\begin{align}
g_{\mu\nu} &= -\frac{1}{2}\eta_{\mu\nu}V(u)\\
V(u) &=   v_0+ \sum_{pr}\frac{v_{pr}}{2}\frac{u^pu^r}{\Lambda^2}  
+ \sum_{prs}\frac{v_{prs}}{3!}\frac{u^pu^ru^s }{\Lambda^3}
+ \sum_{prst}\frac{v_{prst}}{4!}\frac{u^pu^ru^s u^t }{\Lambda^4}
+ \sum_{prstw}\frac{v_{prstw}}{5!}\frac{u^pu^r u^su^tu^w}{\Lambda^5} + \dots
\\
g_{11}(u) &= a_0 +\sum_p a_p \frac{u^p}{\Lambda} + 
\sum_{pr}\frac{a_{pr}}{2}\frac{u^pu^r}{\Lambda^2}  
+ \sum_{prs}\frac{a_{prs}}{3!}\frac{u^pu^ru^s }{\Lambda^3}
 + \dots
 \label{eq.F11}
 \\
g_{22}(u) &= b_0 +\sum_p b_p \frac{u^p}{\Lambda} + 
\sum_{pr}\frac{b_{pr}}{2}\frac{u^pu^r}{\Lambda^2}  
+ \sum_{prs}\frac{b_{prs}}{3!}\frac{u^pu^ru^s }{\Lambda^3}
 + \dots
 \\
g_{12}(u) &= c_0 +\sum_p c_p \frac{u^p}{\Lambda} + 
\sum_{pr}\frac{c_{pr}}{2}\frac{u^pu^r}{\Lambda^2}  
+ \sum_{prs}\frac{c_{prs}}{3!}\frac{u^pu^ru^s }{\Lambda^3}
 + \dots
\end{align}
where the indices $p,r,s,t,w$ take values in $\{1,2\}$, all $a,b,c,v$ parameters are dimensionless, and the ellipses stand for terms inducing 6- or higher-point interactions. The notation generalizes in an obvious way to terms with arbitrarily many $u$ insertions. 

\newpage
For a simple illustration, we restrict our attention to the particular theory with 
\begin{align}
g_{11}(u) &=  1+a_2\dfrac{\phi_2}{\Lambda} 
&
g_{22}(u) &= 1
&
g_{12}(u) &=0\,,
\end{align}
and
\begin{align}
V(u) &= 1+ \frac{v_{11}}{2}\frac{(u^1)^2}{\Lambda^2} + \frac{v_{22}}{2}\frac{(u^2)^2}{\Lambda^2}+
\frac{v_{111}}{3!}\frac{(u^1)^3}{\Lambda^3}
+ \frac{v_{1111}}{4!}\frac{(u^1)^4 }{\Lambda^4}
+ \frac{v_{1112}}{3!}\frac{(u^1)^3(u^2) }{\Lambda^4}
+ \frac{v_{1122}}{2!2!}\frac{(u^1)^2(u^2)^2 }{\Lambda^4}
\nonumber\\
&\quad
+ \frac{v_{11111}}{5!}\frac{(u^1)^5}{\Lambda^5} 
+ \frac{v_{11112}}{4!}\frac{(u^1)^4 u^2}{\Lambda^5} 
+ \frac{v_{11122}}{2!3!}\frac{(u^1)^3 (u^2)^2}{\Lambda^5} 
\end{align}
which leads to the Lagrangian 
\begin{align}
\Lag &=\frac{1}{2}\de_\mu \phi_1 \de^\mu\phi_1\left[1+a_2\frac{\phi_2}{\Lambda}\right]  + \frac{1}{2}\de_\mu\phi_2 \de^\mu\phi_2
-\frac{\Lambda^2 v_{11}}{2} \phi_1^2 - \frac{\Lambda^2 v_{22}}{2}\phi_2^2 - \frac{\Lambda v_{111}}{6}\phi_1^3
\nonumber\\
&
- \frac{v_{1111}}{4!}\,\phi_1^4 
- \frac{v_{1112}}{3!}\,\phi_1^3\phi_2 
- \frac{v_{1122}}{2!2!}\,\phi_1^2\phi_2^2 
- \frac{v_{11111}}{5!}\frac{\phi_1^5}{\Lambda} 
- \frac{v_{11112}}{4!}\frac{\phi_1^4 \phi_2}{\Lambda} 
- \frac{v_{11122}}{2!3!}\frac{\phi_1^3 \phi_2^2}{\Lambda}
\,.
\label{eq.Lag_example}
\end{align}
In this theory, mixings are absent and the two $\phi_1,\phi_2$ scalars are canonically-normalized mass eigenstates, with masses  $M_i^2 = v_{ii}\, \Lambda^2 = - \Lambda^4 \tensor{\bar R}{^\mu_i_\mu_i}/2$. The propagators (Eq.~\eqref{eq.prop_FR})  therefore take the form \begin{equation}
\Delta^{ij}(s)  =\frac{\delta^{ij}}{\Lambda^2} \,\frac{i}{s/\Lambda^2 - v_{ii} }
\end{equation}
In principle, the vacuum of the theory should be determined by solving $\de_i V(u)\equiv 0$. 
For simplicity, we expand around the vacuum solution $u^i=0$.

\subsubsection*{Covariant Feynman rules}

The covariant Feynman rules can be computed by evaluating the relevant covariant tensors appearing in the expressions for $\mathcal{R}_{3,4,5}$ at the vacuum. We performed this operation with the {\tt xTensor} and {\tt xCoba} packages from the {\tt xAct} suite in {\tt Mathematica}~\cite{xAct}. From the Lagrangian in Eq.~\eqref{eq.Lag_example} we obtain the following non-vanishing rules:
\begin{align}
\label{eq.FR_example_first}
 \mathcal{R}_{111} &=  -i\Lambda v_{111}
 \\
 \mathcal{R}_{112} &= \frac{i\Lambda}{2}\,a_2\,\left(2v_{11}-v_{22}\right) 
 \\
 \mathcal{R}_{1111} &= 
 -i v_{1111} + i a_2^2 \left(2v_{11}-\frac{3v_{22}}{4}\right) 
 \\
 \mathcal{R}_{1112} &= -i v_{1112} + \frac{3i}{2}a_2 v_{111} 
 \\
 \mathcal{R}_{1122} &=-iv_{1122} + \frac{ia_2^2}{4}\left(\frac{s_{12}}{\Lambda^2}-8v_{11}+2v_{22}\right)
 \\
 \mathcal{R}_{11111} &= -\frac{iv_{11111}}{\Lambda} -\frac{5ia_2}{\Lambda}v_{1112}+\frac{5ia_2^2}{\Lambda}v_{111}
  \\
 \mathcal{R}_{11112} &= -\frac{iv_{11112}}{\Lambda} + \frac{ia_2}{\Lambda}(2v_{1111}-3v_{1122})
 +\frac{ia_2^3}{2\Lambda}(3v_{22}-10v_{11})
 \\
 \mathcal{R}_{11122} &=
 -\frac{iv_{11122}}{\Lambda} + \frac{3ia_2}{\Lambda}v_{1112}
 - \frac{7ia_2^2}{2\Lambda}v_{111}
 \\
 \mathcal{R}_{11222} &= 
 \frac{3ia_2}{\Lambda}v_{1122}
 -\frac{ia_2^3}{4\Lambda}\left(\frac{2s_{12}}{\Lambda^2}-22v_{11}+5v_{22}\right)
 \label{eq.FR_example_last}
\end{align}
In these expressions $s_{ij} = (p_i+p_j)^2$ where $p_i$ is the incoming momentum of the $i$-th Feynman rule leg. For instance, in $\mathcal{R}_{1122}$, $s_{12}$ is the invariant mass of the two incoming $\phi_1$'s.

It is worth noting  that these expressions depend on both 2- and 0-derivative interactions, which are associated to $a$ and $v$ parameters respectively. This property follows naturally from these objects being vertical tensors on the fibre bundle $E$, and it is crucial in order to account properly for the invariance under field redefinitions.

\subsubsection*{Invariance under field redefinitions}

By definition, the covariant Feynman rules transform as tensors under diffeomorphism field redefinitions. For fixed external indices their expressions are invariant. 
This property can be straightforwardly tested in explicit examples. 
Consider for instance the field redefinition
\begin{equation}
    \phi_1\mapsto \phi_1 + \alpha \phi_1^2\,,
    \qquad \phi_2\mapsto\phi_2\,.
\end{equation}
The action of this shift on the Lagrangian, Eq.~\eqref{eq.Lag_example}, is equivalent to a redefinition of the $a,v$ coefficients: the transformed Lagrangian up to 5-point interactions will have the same polynomial form as the initial one, upon replacing
\begin{align}
\label{eq.example_shift_first}
v_{111} &\to v_{111}+6\alpha\, v_{11}
&
v_{1111} &\to v_{1111} + 12\alpha\,(v_{111}+\alpha\, v_{11})
\\
v_{11111} &\to v_{11111} + 20\alpha\,(v_{1111}+3\alpha\, v_{111})
&
v_{11112} &\to v_{11112} + 12\alpha\, v_{1112}
\\
v_{11122} &\to v_{11122} + 6\alpha\, v_{1122}
&
\end{align}
and re-introducing the four terms 
\begin{align}
\label{eq.example_shift_last}
 a_1 &\to 4\alpha
 &
 a_{11} &\to 8\alpha^2
 &
 a_{12} &\to 4\alpha\, a_2
 &
 a_{112} &\to 8\alpha^2\, a_2\,,
\end{align}
that were taken to vanish in Eq.~\eqref{eq.Lag_example}. These terms would have given the following contributions to the covariant Feynman rules:
\begin{align}
 \Delta\mathcal{R}_{111} &= 
 \frac{3i\Lambda}{2}a_1 v_{11}
 \\
 \Delta\mathcal{R}_{1111} &= 3ia_1 v_{111}
 -\frac{19i}{4}a_1^2 v_{11}
 +2ia_{11}v_{11}
 \\
 \Delta\mathcal{R}_{1112} &=  \frac{ia_1a_2}{4}\left(2v_{22}-15v_{11}\right) + \frac{ia_{12}}{2}\left(3v_{11}-v_{22}\right)
 \\
 \Delta\mathcal{R}_{11111} &=
  \frac{5ia_1}{\Lambda}\left( v_{1111} - \frac{a_2^2}{4}(13v_{11}-2v_{22})
  -\frac{15}{4} a_{11}v_{11}\right)
   +\frac{5i}{\Lambda}a_{11}v_{111}
   +\frac{5i}{4\Lambda}a_1^2\left(-11 v_{111}
  +18 a_1v_{11}\right)
  \\
  &
  +\frac{5i}{4\Lambda}a_{12}a_2(7v_{11}-2v_{22})
 \nonumber
  \\
 \Delta\mathcal{R}_{11112} &= \frac{3ia_1}{\Lambda}\left(v_{1112}-3a_2 v_{111}\right)
 +\frac{ia_{112}}{2\Lambda}(4v_{11}-v_{22})
 +\frac{ia_1^2a_2}{4\Lambda}(76v_{11}-5v_{22}) + \frac{ia_{11}}{2\Lambda}a_2(v_{22}-12v_{11})
 \nonumber\\
 & + \frac{3ia_{12}}{\Lambda}v_{111} + \frac{ia_1a_{12}}{4\Lambda}(5v_{22}-38v_{11})
 \\
 \Delta\mathcal{R}_{11122} &=
 \frac{3i}{4}a_1\left(2v_{1122}
 + 7 a_2^2 v_{11}\right)
 +\frac{ia_2}{2\Lambda}(a_{12}-a_1a_2)\left(\frac{s_{45}}{\Lambda^2}-15v_{11}+3v_{22}\right) 
\end{align}
One can easily check that applying the replacements~\eqref{eq.example_shift_first}--\eqref{eq.example_shift_last} onto each $\mathcal{R}+\Delta\mathcal{R}$  gives again the expressions in Eqs.~\eqref{eq.FR_example_first}--\eqref{eq.FR_example_last}, verifying that the covariant Feynman rules are  invariant under the field redefinition considered.
Importantly, in most cases  the invariance is ensured by non trivial cancellations among shifts of $v$  and $a$ parameters.

\end{document}